\documentclass[letterpaper,english,prl,notitlepage,twocolumn]{revtex4-2}
\usepackage{verbatim}
\usepackage{amsmath}
\usepackage{amssymb}
\usepackage{graphicx}
\usepackage{dsfont}
\usepackage{soul}
\usepackage{mathtools}

\usepackage{hyperref}
\hypersetup{
    colorlinks=true,
    linkcolor=blue,
    filecolor=magenta,      
    urlcolor=cyan,
    pdftitle={Overleaf Example},
    pdfpagemode=FullScreen,
    }

\makeatletter

\usepackage{color}
\usepackage[usenames,dvipsnames,svgnames,table]{xcolor}
\usepackage{graphicx}
\usepackage{comment}
\usepackage{ulem}
\definecolor{violet}{rgb}{0.58, 0.0, 0.83}

\newcommand{\gae}{\lower 2pt \hbox{$\,
\buildrel{\scriptstyle >}\over {\scriptstyle \sim}\,$}}
\newcommand{\lae}{\lower 2pt \hbox{$\,
\buildrel{\scriptstyle <}\over {\scriptstyle \sim}\,$}}

\makeatother

\begin{document}
\title{Testing Models of `Gravity' Using a Quantum Computer}
\author{Christopher I. Timms} 
\affiliation{Ph.D., The University of Texas at Dallas, Richardson, TX, USA} 

\begin{abstract}
This article begins by putting forth a model that shows how the storage and retrieval of information on a wave function that involves quantum entanglement behaves similarly to the concepts of length contraction and time dilation, respectively. An exploration is then made to see if another model can be generated based on the one previously mentioned that guides the time evolution of a quantum system in a manner similar to that of gravity. The answer is made in the affirmative, after testing a series of models, by producing a field that is mediated solely by the transfer of quantum information using both quantum entanglement and wave function collapse. While it is readily acknowledged that the effective field produced may not be gravity, the study provides arguments about why the concepts presented do in fact provide fundamental insights about the true nature of gravity as opposed to merely generating similar behavior to gravity. The study moves on to study how a system with uniform effective potential behaves similarly to dark energy, albeit with a system whose construction is severely limited by computational resources. Finally, an exploration is made as to whether a negative gravitational mass can be produced. It is found that while an isolated particle with negative gravitational mass proves to be problematic according to this model, it seems as though a local region with negative gravitational mass can be produced as long as the resulting fields are cancelled out at larger distances.
\end{abstract}
\maketitle

\section*{Introduction}

There are a wide variety of phases of matter that would be very useful to science and technology, which have proven too unstable to realize in an effective manner. These include various superfluid~\cite{Suthar2021,Konabe2006,Iigaya2006,Diener2007} and superconducting~\cite{Ishida2020,Ginzburg1958,Holczer1991} phases of matter, various phases that exist within Floquet systems~\cite{Rudner2020,Ikeda2021,Harper2020,Moessner2017,Prosen1998,Lazarides2014,Bukov2015,D’Alessio2014}, phases of matter that enable the generation of the $\pi$ Majorana mode~\cite{Karzig2013,Pan2021,Karzig2015,Matthies2022,Liu2019}, or, simply, the orientation of qubits in the spin down state for the initialization of a quantum computer~\cite{Dasgupta2021,Ladd2010,Tacchino2020}. In addition, various mechanisms can benefit from enhanced efficiency to be more practical in the real world. These include various chemical~\cite{Peng2014,Dai2002} and fusion~\cite{Horvath2016,Badziak2012} reactions, photons being transformed into current within a solar cell~\cite{Jennings2011,Lin2020,Stolterfoht2017,Wibowo2020}, or carbon dioxide being recaptured from the atmosphere~\cite{Wilberforce2019,Salvi2019}. This study achieves the production of an effective field, which is mediated through nothing more than the transfer of quantum information, that allows for all states of a particular system to spontaneously evolve towards a particular desired state. While using this force to enhance some of the processes above might be ambitious, the study of this effective force can lead to unknown positive outcomes down the line.

This article begins by putting forward a basic model of time dilation and length contraction~\cite{Einstein2003,Thorne2000} through exploring the way that information is stored and retrieved using entangled qubits on a quantum computer. The article then moves on to exploring whether a force similar to the gravitational force can be achieved on a quantum computer using this model. After a series of models are tested, it is found that when a certain portion of a quantum system is subject both to quantum entanglement operations involving external qubits as well as wave function collapse operations at a higher level than another portion of the quantum system, particles occupying the system will reorient towards the regions where the operations are more likely to occur. Various arguments are given throughout this paper for why this is in fact the gravitational force and not just similar behavior to the gravitational force, but it is also acknowledge that there is still some room for further proof.

This model is then used to enhance diffusion of the system by applying the entanglement and wave function collapse operations with equal probability throughout the system, which is supposed to reproduce similar behavior to dark energy~\cite{Amendola2010}. Finally, a setup is explored where the field produced points away in two directions from a central portion of a quantum system; thereby modelling antigravity and negative gravitational mass~\cite{Bondi1957,Bonnor1989,Bonnor1964}. It is found that, while producing a standalone negative gravitational mass is problematic, a local region with negative gravitational mass can be produced if regions of positive gravitational mass cancel out the corresponding fields over larger distances.

\section*{Methods}

\subsection*{Preliminary Motivations}

The original incentive for writing this paper concerned a theory given by the author whose intention was originally to compress the amount of RAM needed to perform a quantum computational operation of a reasonable size and therefore place the problem on a classical computer~\cite{Jozsa1997}. What was found was that only the amount of RAM for a series of single qubit operations could be compressed, but when multiple qubit operations were involved, it was not clear how the compression of the RAM could be achieved. The `compression' of single qubit operations is a rather simple process involving a set of control qubits as well as a single target qubit. If there is a single control qubit, then there are two `compressed' qubits that can be stored on the target qubit; with these two qubits being $\Psi_1^T=[a\ b]$ and $\Psi_2^T=[c\ d]$. The transpose is used as a method to save space. If a single control qubit is involved along with the target qubit, then the wave function then becomes:

\begin{equation*}
\Psi_{\mathrm{tot}}^T = \mathrm{CU}(\mathrm{C}_1\otimes \mathrm{T}_1) =[\Psi_1^T\ \Psi_2^T] = [a\ b\ c\ d].
\end{equation*}
where $\mathrm{C}_1$ is the control qubit, $\mathrm{T}_1$ is the target qubit, and $\mathrm{CU}$ is a controlled unitary that encodes the appropriate information on the wave function.

Meanwhile, if two control qubits are involved, the corresponding wave function becomes:

\begin{multline*}
\Psi_{\mathrm{tot}}^T = \mathrm{CU}(\mathrm{C}_1\otimes \mathrm{C}_2\otimes \mathrm{T}_1) = [\Psi_1^T\ \Psi_2^T\ \Psi_3^T\ \Psi_4^T] =\\
[a\ b\ c\ d\ e\ f\ g\ h].
\end{multline*}
where $\Psi_1^T = [a\ b]$, $\Psi_2^T = [c\ d]$, $\Psi_3^T = [e\ f]$, $\Psi_4^T = [g\ h]$, $\mathrm{C}_1$ is the first control qubit, $\mathrm{C}_2$ is the second control qubit, and $\mathrm{T}_1$ is the target qubit. As can be seen, if $n$ describes the number of control qubits, then the number of unentangled qubits that can be `compressed' onto the wave function becomes $2^n$. This allows for the space required for single qubit operations to be reduced logarithmically.

However, it would be substantially more impactful to compress the space required for multi-qubit operations~\cite{Jozsa1997}. The ket $\Psi_\mathrm{tot} = [a\ b\ c\ d]$ is not easy to express entanglement with because it does not seem readily apparent to have the results $a$ and $b$ depend on $c$ and $d$ or vice versa. So instead, the attempt is to involve two copies of $\Psi_{\mathrm{tot}}$, such that $\Psi_{\mathrm{ult}}^T = \Psi_{\mathrm{tot}}\otimes \Psi_{\mathrm{tot}}$, which becomes:
\begin{equation*}
\Psi_\mathrm{ult}^T = [aa\ ab\ ac\ ad\ ba\ bb\ bc\ bd\ ca\ cb\ cc\ cd\ da\ db\ dc\ dd]
\end{equation*}
Now, the entanglement operation is performed such that if the outcome of any of the two sub-kets $\Psi_{\mathrm{tot}}$ is $a$, then the outcomes $c$ or $d$ are left alone, but if the outcome of any of the two sub-kets $\Psi_{\mathrm{tot}}$ is $b$, then the outcomes $c$ and $d$ are switched. In this case, the ket $\Psi_{\mathrm{ult}}$ becomes:
\begin{equation*}
\Psi_\mathrm{ult}^T = [aa\ ab\ ac\ ad\ ba\ bb\ bd\ bc\ ca\ db\ cc\ cd\ da\ cb\ dc\ dd]
\end{equation*}
as can be seen, fewer of these components involve the information relating to entanglement compared to:
\begin{equation*}
\Psi_{\mathrm{ult},2}^T = \Psi_1^T\otimes \Psi_2^T = [ac\ ad\ bc\ bd]
\end{equation*}
where the entangled ket would become $\Psi_{\mathrm{ult},2}^T=[ac\ ad\ bd\ bc]$.
The problem only grows worse, if the compression of a larger number of qubits is attempted.

The two lines of logic behind having the `compression' of single qubit operations be analogous to relativity and the curvature of spacetime are that the `compression' of entangled qubits would produce a kind of technological singularity, which can be seen as analogous to the singularity at the center of a black hole~\cite{Eden2013,Modis2022}, and if the number of compressed single qubits increases by a factor of two, then the number of shots needed to obtain the same degree of accuracy for describing the system also increases by a factor of around two. So, according to this model, the degree to which qubits are compressed into a wave function can be seen as analogous to the contraction of space and the number of shots required to obtain a certain degree of accuracy can be seen as analogous to the dilation of time. Therefore, if both of these assumptions hold, then it seems that the presence of entanglement serves as a barrier to the achievement of the actual singularity; otherwise information would not be conserved and causality would be violated~\cite{Herbert1982,Dieks1982,Wootters1982,KumarPati2000,Ghirardi2013}. It would be phenomenal to achieve the singularity so that strides could be made in terms of informational processing~\cite{Jozsa1997}, but it seems that the prevention of a singularity is more important from the universe's point of view. This does, however, provide a method for a quantum internet that can provide high data rates as well as high degrees of encryption as seen in this \href{https://github.com/htim327/QuantumGravity/blob/main/docs/source/PreliminaryMotiviations.rst}{GitHub tutorial}.

\subsection*{Basic Setup of Where `Gravity' is Tested}
\begin{figure}[t]
    \includegraphics[width=\columnwidth]{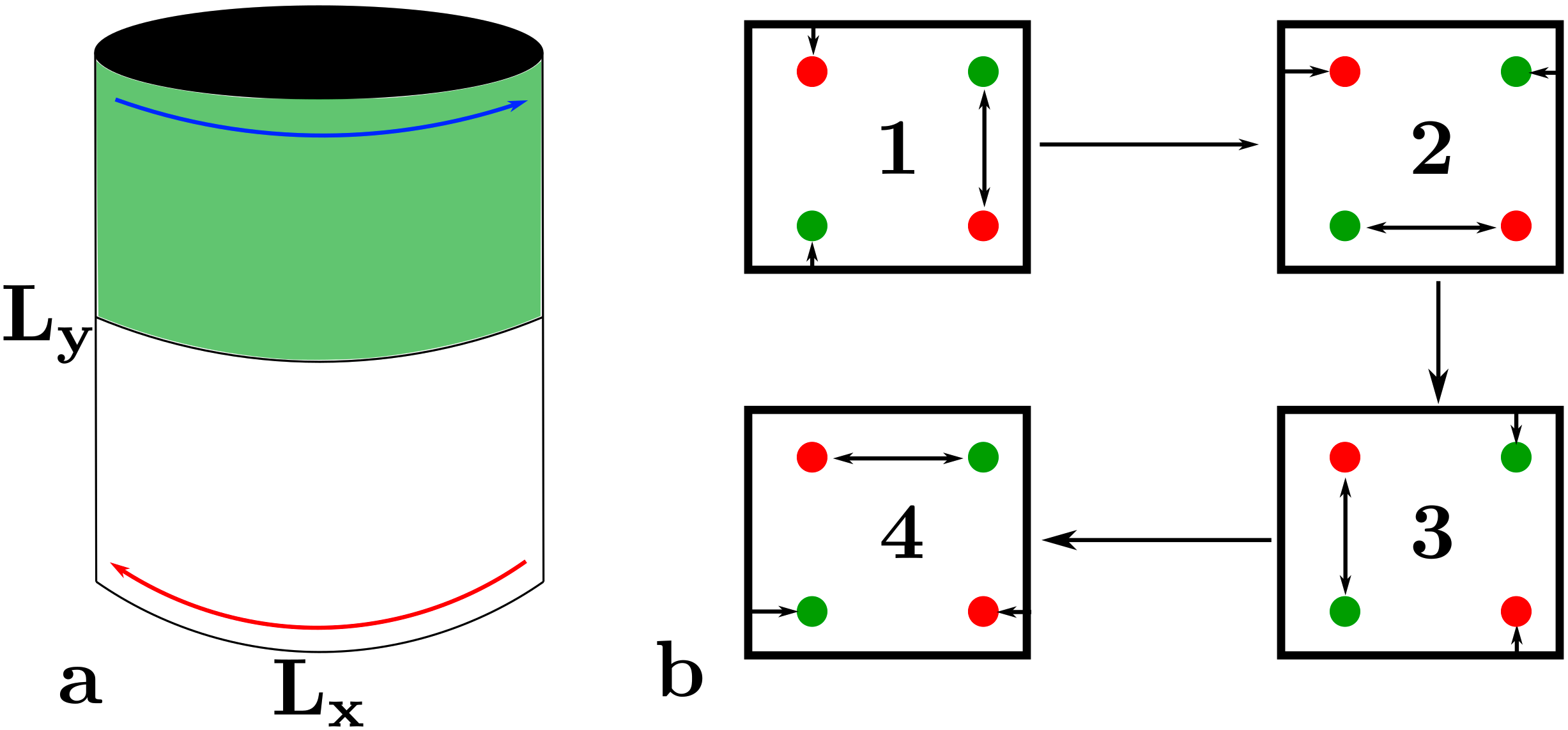}
    \caption{This figure presents the basic setup for the AFAI lattice. b) represents the first four driving steps of the Floquet drive, where the green sites represent the A sites and the red sites represent the B sites. The boxes that bound these sites are intended to show that there other sites that compose this lattice other than the two A and two B sites. The arrows show how the particles move during each of the driving steps with the number in the center defining the driving steps of interest. For the first driving step, particles move between the A and B sites at the right of the sub-diagram, whereas the particles that exist at the B site on the left move up one site and the particles that exist at the A site on the left move down one site as shown by the arrows that point out of the sub-figure. Particles on the left side of this diagram will also move to the B site from one site above the B site and to the A site from one site below the A site. If implemented perfectly and not along the boundaries of the lattice, this will allow a particle that starts at a particular site to return back to its initial position. This does not show the final driving step that is used to implement a chemical potential. a) shows how the driving of the two dimensional cylindrical lattice produces counter-propogating edge state currents at the top and bottom of the cylinder.}
    \label{AFAISetup}
\end{figure}

Figure~\ref{AFAISetup} shows the setup that is used to test the various models of `quantum gravity.' This setup is exactly the same as that of the Anomalous Floquet-Anderson Insulator, which uses a five step Floquet drive of a two dimensional cylindrical lattice to produce edge states~\cite{Timms2021,Titum2016,Kundu2020}. The first four driving steps produce cyclotron-like orbits in the bulk if implemented perfectly, such that the particle returns back to its initial position after these steps are done. The first four driving steps are illustrated in Figure~\ref{AFAISetup} b) and is realized using the Hamiltonian $H_n=-J\sum_{<ij>_n} c_i^\dagger c_j$, where $n$ is the driving step number and $i$ and $j$ are the sites that the particles hop between. The purpose of the fifth step is to generate a chemical potential and uses the Hamiltonian $H_5=\Delta \sum_{i} \eta_i c_i^\dagger c_i$, where $\eta_i=+1$ ($-1$) for the A (B) sites, which will be defined further on. The time for each Floquet cycle, where each of the five driving steps are applied, is given by $T=2\pi/\Omega$, $J$ and $\Delta$ are chosen such that $J=5\Omega/4$ and $\Delta = 0.4$, and $\Omega=\hbar=1$~\cite{Timms2021}. To understand the AFAI more in depth, see the paper by Timms et al.~\cite{Timms2021} as well as this page on \href{https://github.com/htim327/NonHermitianDriving/blob/main/docs/source/AFAI.rst}{GitHub}.

Figure~\ref{AFAISetup} a) is intended to show that particles present at the upper or lower edges of the two dimensional cylindrical lattice will produce counter-propagating edge state currents if the Floquet drive is implemented reasonably well~\cite{Timms2021,Titum2016,Kundu2020}. In addition, the lattice is divided into unit cells that have an A site on the left and a B site on the right in the x-direction, meaning the width of these unit cells in the y-direction is one site. The dimensions of this lattice in terms of unit cells is given by $\mathrm{L}_\mathrm{y}$ in the y-direction and $\mathrm{L}_\mathrm{x}$ in the x-direction~\cite{Timms2021,Titum2016,Kundu2020,RodriguezMena2019}.

Normally, only one half of the cylinder would be populated with particles so that a non-zero Winding number can be measured due to the lack of a counter-propagating current~\cite{Timms2021,Titum2016,Kundu2020}. The green portion of Figure~\ref{AFAISetup} a) depicts the portion of the cylinder that is populated. However, the purpose of this study is not to produce edge state currents, but to rather have particles diffuse throughout the lattice by using a random walk and then guide the particle to become more likely to occupy a certain region of the lattice simply by the transfer of quantum information. To achieve the random walk, temporal disorder is used where if the normal amount of time for each driving step is $T/5$, the implementation of temporal disorder is defined by $T_n = T(1+\delta_n)/5$ where $T_n$ defines the timing for the $n^{th}$ driving step, $\delta_n$ is sampled uniformly with $\delta_n\in [-W_T,W_T]$, and $W_T$ determines the strength of the temporal disorder~\cite{Timms2021}. The strong levels of temporal disorder, where $W_T=0.5$ for this study, will cause the topology of the system to become negligible~\cite{Timms2021}.

\subsection*{Models of `Quantum Gravity' Tested}

\begin{figure}[t]
    \includegraphics[width=\columnwidth]{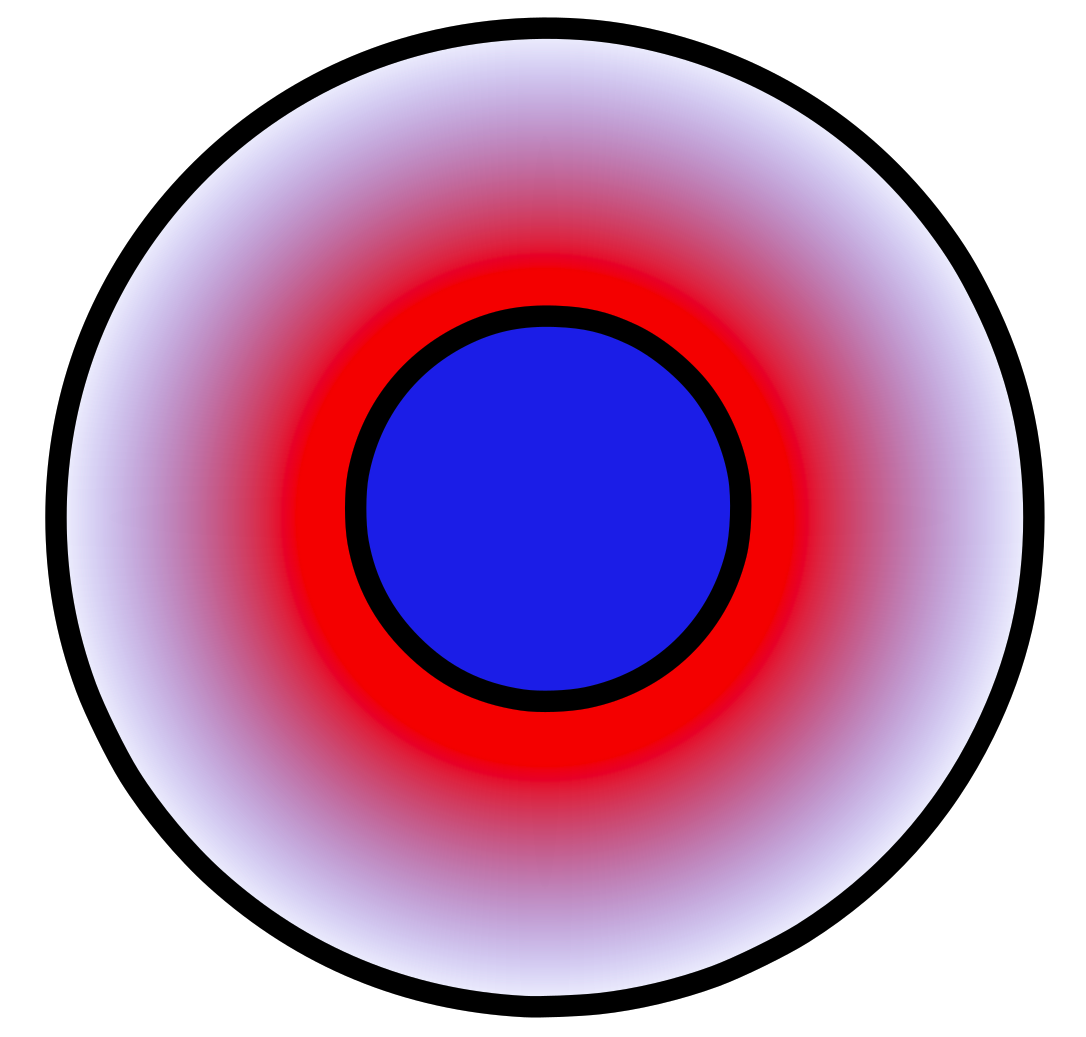}
    \caption{This image depicts a gravitational body in the center that has the color blue. Then there is also space that surrounds this body, which has a gradient that flows from red to purple to white. This gradient is supposed to represent the degree to which the gravitational body is able to obtain information about the surrounding space. Due to the inverse-square law, the body is most likely to gain information about the region close to it in bright red using various real or virtual information carriers. Then as the radius from the body increases, the body is less and less able to obtain information about the various forms of matter present.}
    \label{GravitationalGradient}
\end{figure}

Figure~\ref{GravitationalGradient} illustrates the hypothesis about how the gravitational field works for all of the models tested. There is a central gravitational body in blue and the gravitational potential is shown by the color gradient that flows from red to purple to white. Due to the inverse-square law~\cite{Saslow2002,Brownson2013}, information carriers, whether virtual or real, are able to obtain the most amount of information about the distribution of matter for the area in red and store that information in the matter that composes the gravitational body. As the radius away from the gravitational body increases, less and less information about the matter distribution is obtained by the gravitational body until the area in white is reached where very little information is obtained. One potential reason for why this mechanism would allow for the gravitational force to be linearly dependent on the mass of a smaller body that exists a certain distance away from the gravitational body is that the informational content of a mass confined within a particular region has a maximal value set by the Bekenstein bound~\cite{Bekenstein1981,Bekenstein2005,Bekenstein2008} and so therefore, the maximal amount of information that can be transferred from the body to the gravitational body is linearly dependent on the mass, but this is just a hypothesis.

\subsubsection*{Using Entanglement}

There are three basic models that are tested to see if a `gravitational' force can be produced using a single particle with an undetermined mass that populates the cylindrical lattice. The first model produces a gradient with respect to how much entanglement can be generated during a given time that is dependent upon the location of the particle in the y-direction. This works by dividing the time evolution of each of the driving steps by a certain factor, such that each driving step is composed of a discrete number of time steps. After a unitary evolves the system for each of these time steps, a random value for the y-index of interest is obtained where y-indices with a higher value are more likely to be chosen. This choice is made by drawing a random number and then finding the lowest value of a probability vector similar to $p=[1/1000\ 1/100\ 1/10\ 1]$ that is still higher than the random number drawn; with the position of this lowest value determining the y-index of interest.

Then a process is implemented where all of the possible x-indices and values for $\alpha$ are iterated over given the chosen value for the y-index. For each iteration, an external qubit is added and if the site of interest is populated with a particle, the spin of the external qubit is flipped, then the reduced density matrix of the system is calculated where the external qubit is effectively removed. The control operation that flips the qubit depending on the presence of a particle at the site of interest is given by:

\begin{equation*}
\mathrm{CU} = |x,y,\alpha\rangle \langle x,y,\alpha |\otimes \sigma_x + I_{\sim x,y,\alpha}\otimes I_2
\end{equation*}
where $|x,y,\alpha\rangle$ represents the site of interest, $\sigma_x$ is the Pauli-x matrix, $I_{\sim x,y,\alpha}$ is the identity matrix that represents every site except for the site of interest, and $I_2$ is the $2\times 2$ identity matrix. The code that runs the relevant simulation can be seen with this \href{https://github.com/htim327/QuantumGravity/blob/main/docs/source/EntanglementGradient.rst}{documentation on GitHub}.

\subsubsection*{Using Wave Function Collapse}

The next model tested uses wave function collapse as opposed to quantum entanglement. So as the value of the y-index increases, the probability of performing a measurement on the sites increases. In addition, this method only collapses the wave function such that it is only localized in the y-direction, but not the x-direction. The way that this wave function collapse method works is that, again, the time evolution for each driving step is divided by a certain factor, such that a single driving step takes place after an appropriate number of the time steps generated. After a single time step, a random number determines where the presence of the particle will be measured with respect to its y-index, but not its x-index. 

This process is administered by setting up a vector that determines the probability of performing a measurement of a particular y-index; this vector can be $p=[1/1000\ 1/100\ 1/10\ 1]$ for a system of size $L_y=4$. $p$ is then iterated over from left to right and the y-index is iterated over from 0 to 3. If the random number generated is less than $p(i)$, where $i$ is the iteration number, then the y-index measured corresponds to the current iteration number. Next, the probability of observing the particle at the given y-index is calculated using the appropriate density matrix.

The wave function is then partially collapsed by taking the appropriate density matrix and then calculating the corresponding eigenvalues $E_i$ and the eigenvectors $|\Phi_i\rangle$ of this matrix. Another random number is then calculated and if the random number is less than the probability of observing the particle at the given y-index, then all of the components of all of the eigenvectors that do not correspond to the given y-index are set to zero, otherwise only the components that correspond to the given y-index are set to zero. These eigenvectors are then normalize, yielding $|\Phi_{i,f}\rangle$. Finally, the density matrix is reconstructed using $\rho_{\mathrm{new}}=\sum_i E_i |\Phi_{i,f}\rangle\langle \Phi_{i,f}|$, with $\rho_{\mathrm{new}}$ being re-normalized again, and the time evolution is allowed to begin again. The code that performs this simulation is posted on this \href{https://github.com/htim327/QuantumGravity/blob/main/docs/source/MeasurementGradient.rst}{GitHub page}.

\subsubsection*{Using Entanglement and Wave Function Collapse}

The final method works by, once again, dividing the time evolution by a certain factor and after the implementation of the unitary for that time step, an iterative process is performed that randomly chooses which sites will be used for the entanglement and wave function collapse operations. The number of iterations for this process was arbitrarily chosen to be equal to the number of sites on the lattice. During each iteration, first, the process is repeated where the y-index for the site that is going to be measured is randomly chosen using a random number and a probability vector similar to $p=[1/1000\ 1/100\ 1/10\ 1]$, after which the x-index and $\alpha$ are randomly chosen using a uniform probability distribution. Then, the probability of a particle occupying this site is calculated and a random number generator determines whether to set the occupation probability of this site to zero or one. 

During the same iteration, another site of the lattice is chosen to become entangled with an external particle. The y-index is again chosen using a probability vector similar to $p=[1/1000\ 1/100\ 1/10\ 1]$ and also the x-index and $\alpha$ are chosen using a uniform probability distribution. An external qubit is then added and if a particle is present at the site of interest, the external qubit is flipped from the spin down state to the spin up state, otherwise the external qubit is left alone. The external qubit is then separated from the system through calculating the reduced density matrix. After this, the next iteration proceeds and after all of the iterations, the unitary for the next time step is implemented. The code for this simulation is presented on this \href{https://github.com/htim327/QuantumGravity/blob/main/docs/source/MeasurementAndEntanglementGradient.rst}{GitHub page}.

\section*{Results}

\subsection*{Models Tested}

\subsubsection*{Using Entanglement}

\begin{figure}[t]
\centering
\includegraphics[width=\columnwidth]{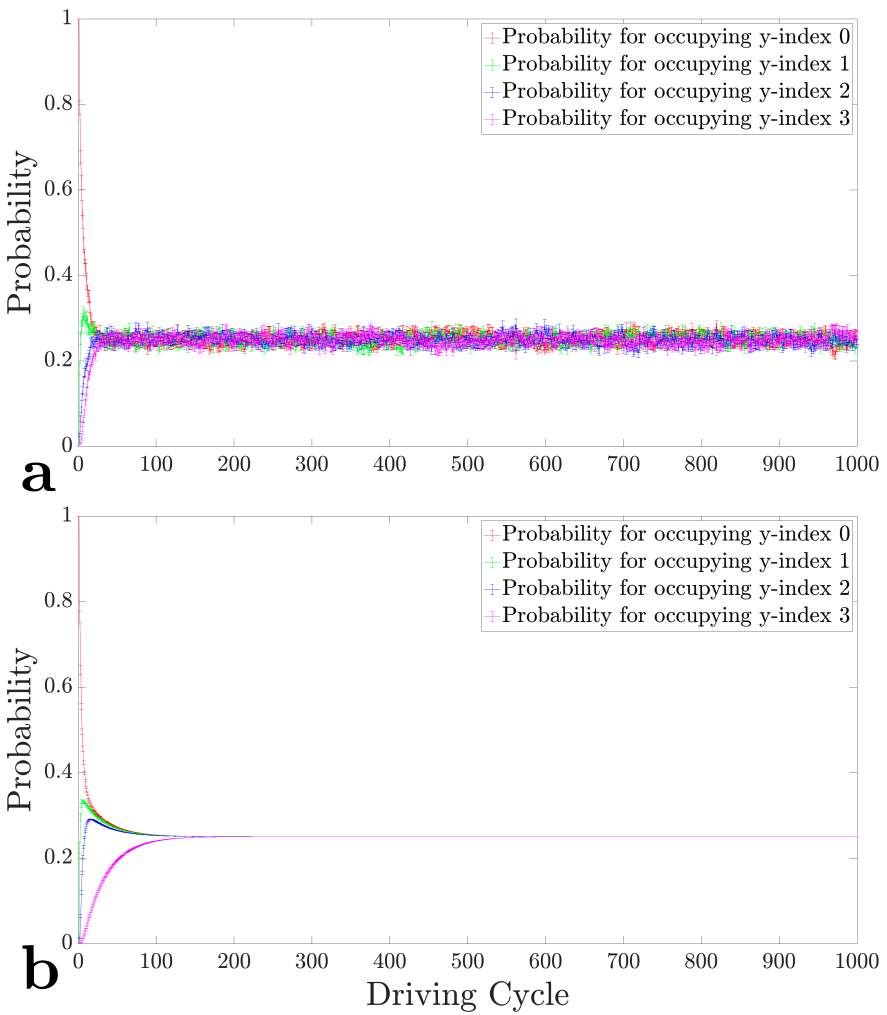}
\caption{a) displays the results for the system that is never involved in entanglement with an external qubit whereas with b), the number of times that a site is entangled with an external qubit is dependent on the value of the y-index; with higher values of the y-index being associated with a greater frequency of entanglement with an external qubit. For both of these sub-figures, the number of noise realizations is 100, $p=[1/1000\ 1/100\ 1/10\ 1]$, and the y-index is randomly chosen using the probability index $p$ 100 times per driving step. In both cases, the probability for the particle occupying each of the y-indices asymptotes to 0.25, which means that an effective force was not produced.}
\label{EntanglementGradient}
\end{figure}

Figure~\ref{EntanglementGradient} displays the results for the model where the sites are entangled with an external qubit, such that if the particle occupies the site of interest, the external qubit is flipped from the spin down state to the spin up state. In addition, the degree to which this site dependent entanglement is produced is determined by the value of the y-index where the greater the value of the y-index, the greater the frequency with which the particle occupying that y-index becomes entangled with an external qubit. The hope was that the particle would travel the path that would generate the most amount of entropy~\cite{Wei2003,Munro2001}, which would mean that the particle would `gravitate' to having higher values of the y-index. However, Figure~\ref{EntanglementGradient} b) clearly shows that the particle has an equal probability of occupying each of the y-indices at longer time scales when this gradient of entropy production is present. This behavior is probably related to that of recent studies, which have shown that the way in which this entanglement is implemented is analogous to the presence of chemical potential disorder~\cite{timms2023floquet} where in this case, the chemical potential would have a sort of random time dependence.

\subsubsection*{Using Wave Function Collapse}

\begin{figure}[t]
\centering
\includegraphics[width=\columnwidth]{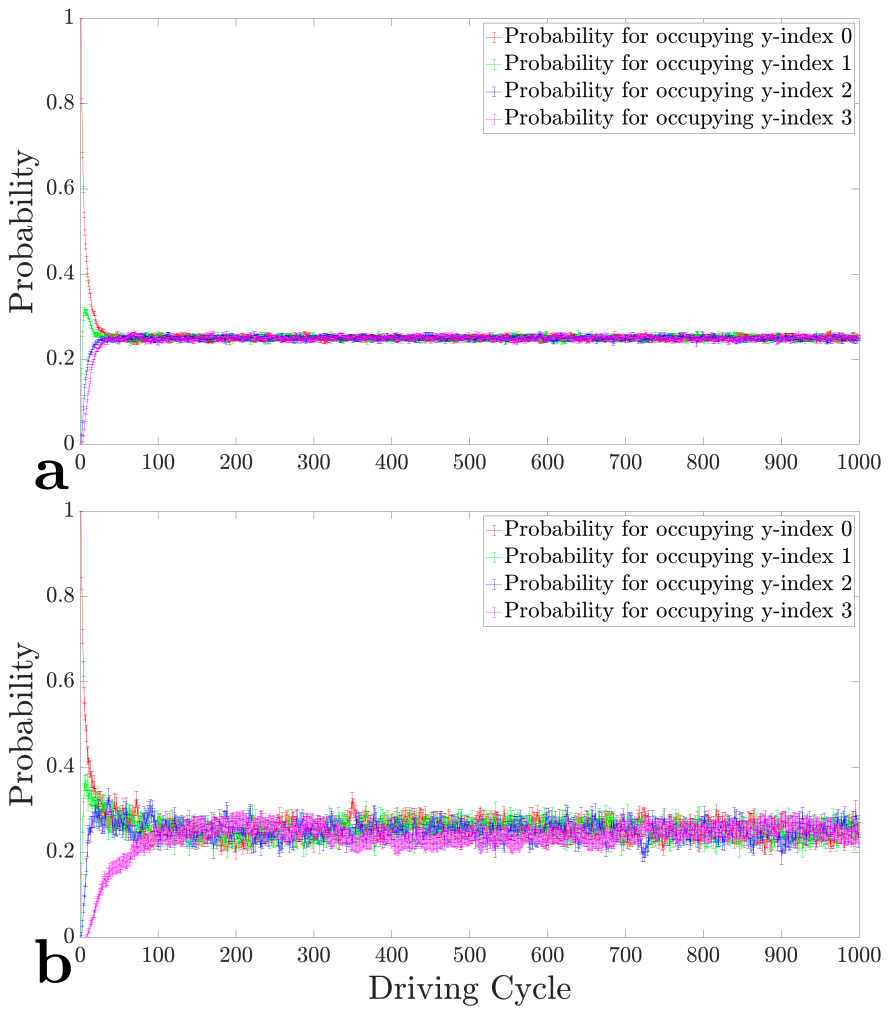}
\caption{a) displays the results for the system where the wave function collapse method is not used whereas with b), the number of times that the particle is measured as being a part of a particular y-index is dependent on the value of the y-index itself; with higher values of the y-index being associated with a higher probability of being measured. The number of noise realizations is 600 for both of these plots, $p=[1/1000\ 1/100\ 1/10\ 1]$, and the y-index is randomly chosen using the probability vector $p$ 100 times per driving step. The probability for the particle occupying each of the y-indices asymptotes to 0.25 for both cases, which means that an effective force was not produced.}
\label{WaveFunctionCollapse}
\end{figure}

Figure~\ref{WaveFunctionCollapse} shows how the system behaves using the wave function collapse method where the particle is more likely to be measured after each of the time steps if it has a higher value of the y-index. The motivation behind this model was that, despite the fact that the process of wave function collapse is not completely understood, wave function collapse could still be partially explained as generating von Neumann entropy and cause the particle to gravitate towards regions where more entropy is produced, while also not involving the localization effects that result from entanglement. The hypothesis that wave function collapse is associated with von Neumann entropy is based on the similar behavior of a collapsed wave function and a wave function that has been entangled with another wave function~\cite{Wallace2012}. However, both the plot that does not involve the wave function collapse method a) and the plot that does involve collapses of the wave function b) show that the probability of the particle occupying each of the sites asymptotes to 0.25 for all values of the y-index, which means that an effective force has not been generated. This could be due to the quantum Zeno effect~\cite{Itano1990,Koshino2005} preventing the particle from both being likely to transition to high value y-indices and from being likely to transition away from high value y-indices with equal probability.

\subsubsection*{Using Entanglement and Wave Function Collapse}

Figure~\ref{Gravity} presents the results for the case where a random site is chosen for the wave function collapse operation and then independently, another random site is chosen again for the entanglement operation during an iterative process that occurs after the system has been evolved for each time step. Again, these time steps are produced when the time evolution of each driving step is divided by a certain factor. These plots show the probability of the particle occupying y-indices 0, 1, 2, and 3 for sub-figures a), b), c), and d), respectively, with the probability vector being $p=[1/1000\ 1/100\ 1/10\ 1]$ for both the the wave function collapse and entanglement operations. It is clearly shown that at longer time scales, the particle is more likely to occupy higher y-indices and for the most part, the difference as indicated by the error bars is greater than $5\sigma$ for y-indices 0 and 3.

These figures demonstrate that an effective force is being produced by the transfer of quantum information through the processes of quantum entanglement and wave function collapse. It's reasonable that the effective force produced requires the use of wave function collapse in addition to quantum entanglement in light of recent studies that have shown that the implementation of quantum entanglement involving each of the sites results in behavior similar to that of chemical potential disorder~\cite{timms2023floquet}. The wave function collapse, therefore, could allow for the disorder generated by the entanglement operations to cause the particle to diffuse while reducing the degree to which the wave function of the particle interferes with itself, which would otherwise result in behavior similar to that of Anderson localization~\cite{Flores2004}. The high frequency of wave function collapse for higher values of the y-index could also serve to localize the particles at these higher values of the y-index due to the quantum Zeno effect~\cite{Itano1990,Koshino2005}.

However, the required presence of wave function collapse for the existence of this effective force makes it difficult to reconcile this model with that described in the sub-section `Preliminary Motivations' within `Methods.' It could be that the process of wave function collapse is dependent upon various arrangements of entanglement distributions, but this is yet to be determined as the measurement problem has not been solved~\cite{Wallace2012}. Some ways to test if this effective force is in fact gravity is to see how this force interferes with the gravitational waves detected by LIGO or to see if this force can be produced in a manner that can be observed by LIGO.

\subsection*{Dark Energy}

This section tests a system using the same model as that of sub-section titled `Using Entanglement and Wave Function Collapse' except that there is equal probability to randomly choose any of the y-indices on top of the already implemented equal probability of choosing any of x-indices and values of $\alpha$. This is done using the probability vector $p=[0.25\ 0.5\ 0.75\ 1]$ for both the entanglement and wave function collapse operations. This is supposed to model a constant energy density that causes wave function collapse or entanglement for the particle regardless of its location defined by the y-index; with dark energy being associated with a presence of a constant energy density~\cite{Amendola2010}. 

Figure~\ref{DarkEnergy2} illustrates what happens when the time evolution of the driving steps is divided by a factor of two and the entanglement and wave function collapse operations are implemented after each of these time steps. This plot shows that the particle does in fact diffuse faster than the case where the entanglement and wave function collapse operations are not implemented (blue). a) shows that the particle diffuses away from its initial position (y-index 0) and moves to populate b), c), and d) (y-indices 1, 2, and 3, respectively). Meanwhile Figure~\ref{DarkEnergy10} displays the results when each of the driving steps are divided into 10 time steps whereas Figure~\ref{DarkEnergy100} is for when the driving steps are divided into 100 time steps. Both of these plots show that when the constant energy density is increased, the diffusion slows down and the particle becomes more localized, which agrees with what is known about the effects of gravitational time dilation in the presence of high energy densities~\cite{Thorne2000}.

It is difficult to know whether the data presented is actually a representative model of dark energy. It is well known that dark energy expands distances and space between objects within the universe~\cite{Amendola2010}, but the model presented here always contains a finite number of sites that are a fixed distance apart. It could be that this is in fact a good model of dark energy, but the particle is prevented from expanding to more distant locations by the boundary conditions of the system, which would be difficult to construct on a scale that the universe's dark energy acts on.

\subsection*{Anti-Gravity}

\begin{figure}[t]
\centering
\includegraphics[width=\columnwidth]{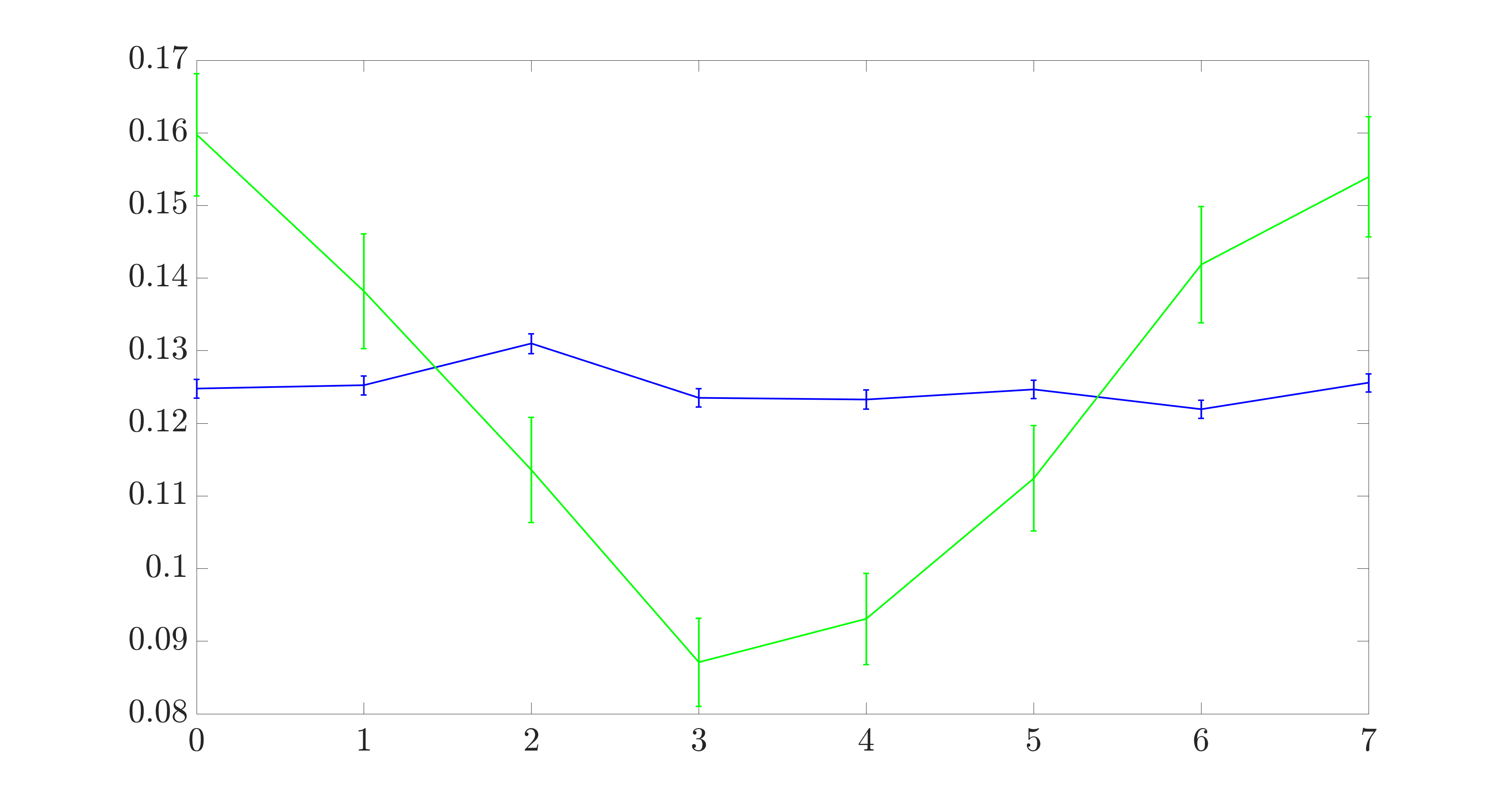}
\caption{This graph uses the model where both wave function collapse and entanglement operations are used. The probability that these operations will be implemented are weakest for y-indices 3 and 4. The probability then steadily grows from y-indices 4 to 7 and from 3 to 0. The x-axis displays the y-index, the y-axis displays the probability of the particle occupying the particular y-index, the green curve represents the system affected by the operations, and the blue curve represents the system unaffected by the operations. The probabilities plotted on this graph are taken after the system has been evolved for 1000 driving cycles and 1903 noise configurations have been used.}
\label{AntiGravityOccupation}
\end{figure}

\begin{figure*}[t]
\centering
\includegraphics[width=1\textwidth]{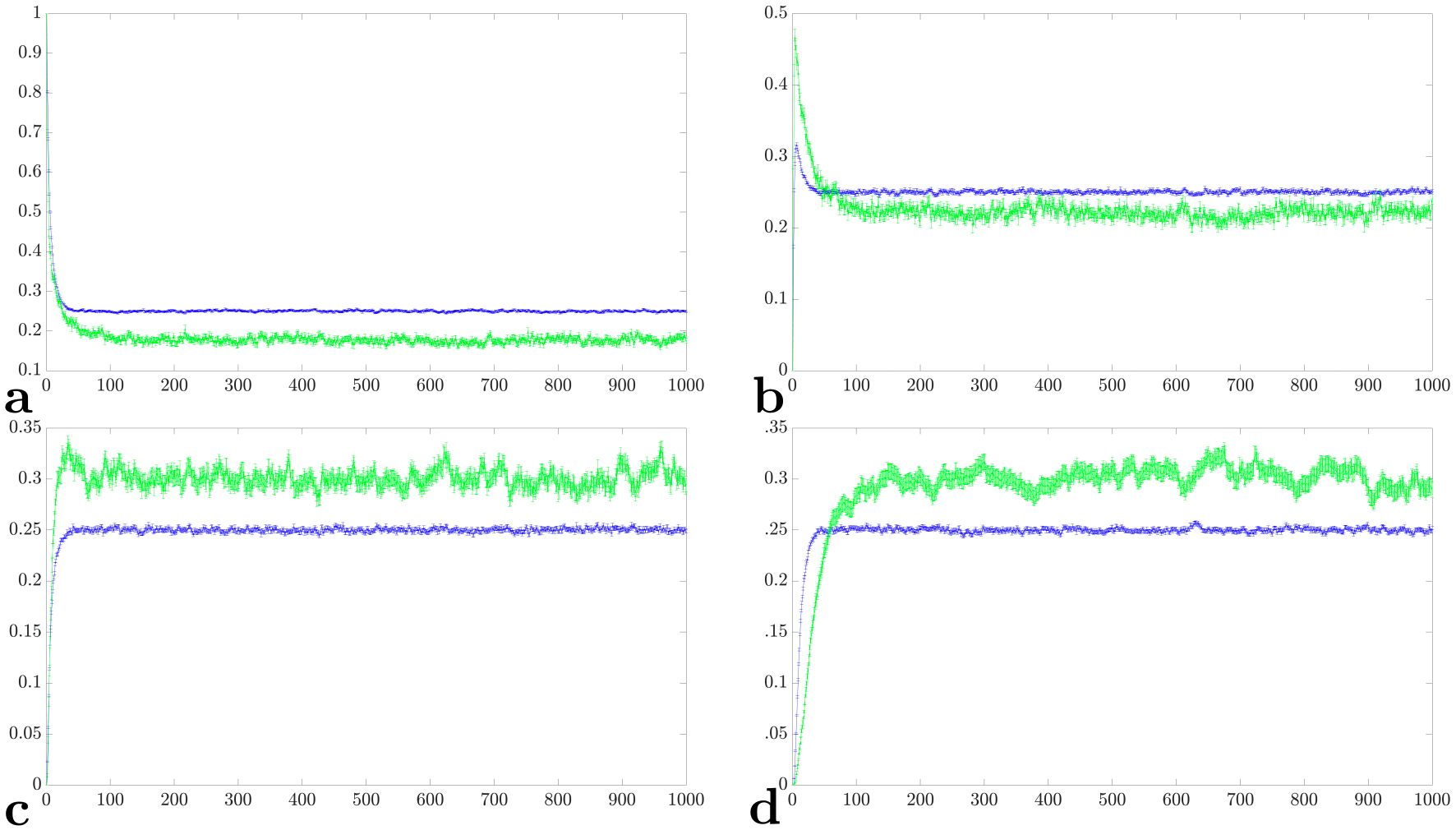}
\caption{These sub-figures present the model where after the system is evolved for each time step, the system undergoes an iterative process and during each iteration, a random site is chosen for the wave function collapse operation and then the entanglement operation. The y-index chosen for these operations have an increasing probability of being chosen for y-indices that have a higher value, while all of the x-indices and all of the values of $\alpha$ have an equal probability of being chosen. The sub-figures of a), b), c), and d) correspond to the probabilities of the particle occupying y-indices 0, 1, 2, and 3, respectively. The particle was initialized in the zeroth y-index and 3675 noise realizations were used for this plot. The system unaffected by the wave function collapse and entanglement operations is plotted in blue and the probability for the particle to occupy each of the y-indices for this case asymptotes to 0.25, but this is not the case for the plots in greeen where the entanglement and wave function collapse operations were used.}
\label{Gravity}
\end{figure*}

\begin{figure*}[t]
\centering
\includegraphics[width=1\textwidth]{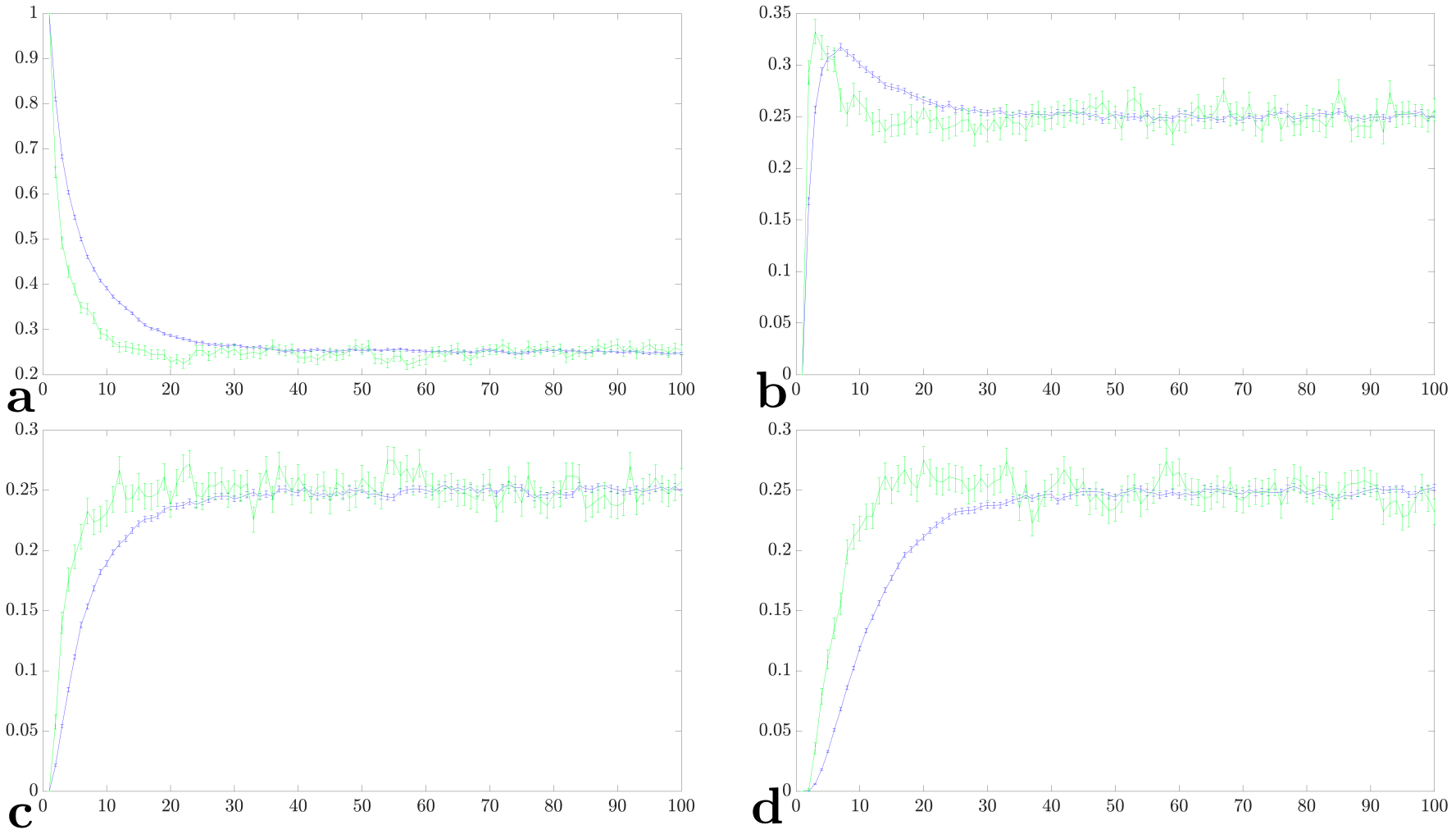}
\caption{This figure displays the results for the case where the probability of a site being acted upon by either the entanglement or wave function collapse operations is independent of either the y-index, x-index, or value for $\alpha$; thus causing the system to have a constant energy density associated with it. The time evolution of each of the driving steps is divided by a factor of two and after each of these time steps, the system undergoes an iterative process where the number of iterations is equal to the number of sites. During each iteration, a random site is chosen to perform the wave function collapse operation and then another random site is chosen to perform the entanglement operation. a) corresponds to the y-index of 0, b) corresponds to the y-index of 1, c) corresponds to the y-index of 2, and d) corresponds to the y-index of 3. The total number of noise realizations is 1600. The blue curve represents the case where the system is unacted upon by the entanglement or wave function collapse operations. The x-axis displays the driving cycle and the y-axis displays the probability of occupying the y-index of interest}
\label{DarkEnergy2}
\end{figure*}

\begin{figure*}[t]
\centering
\includegraphics[width=1\textwidth]{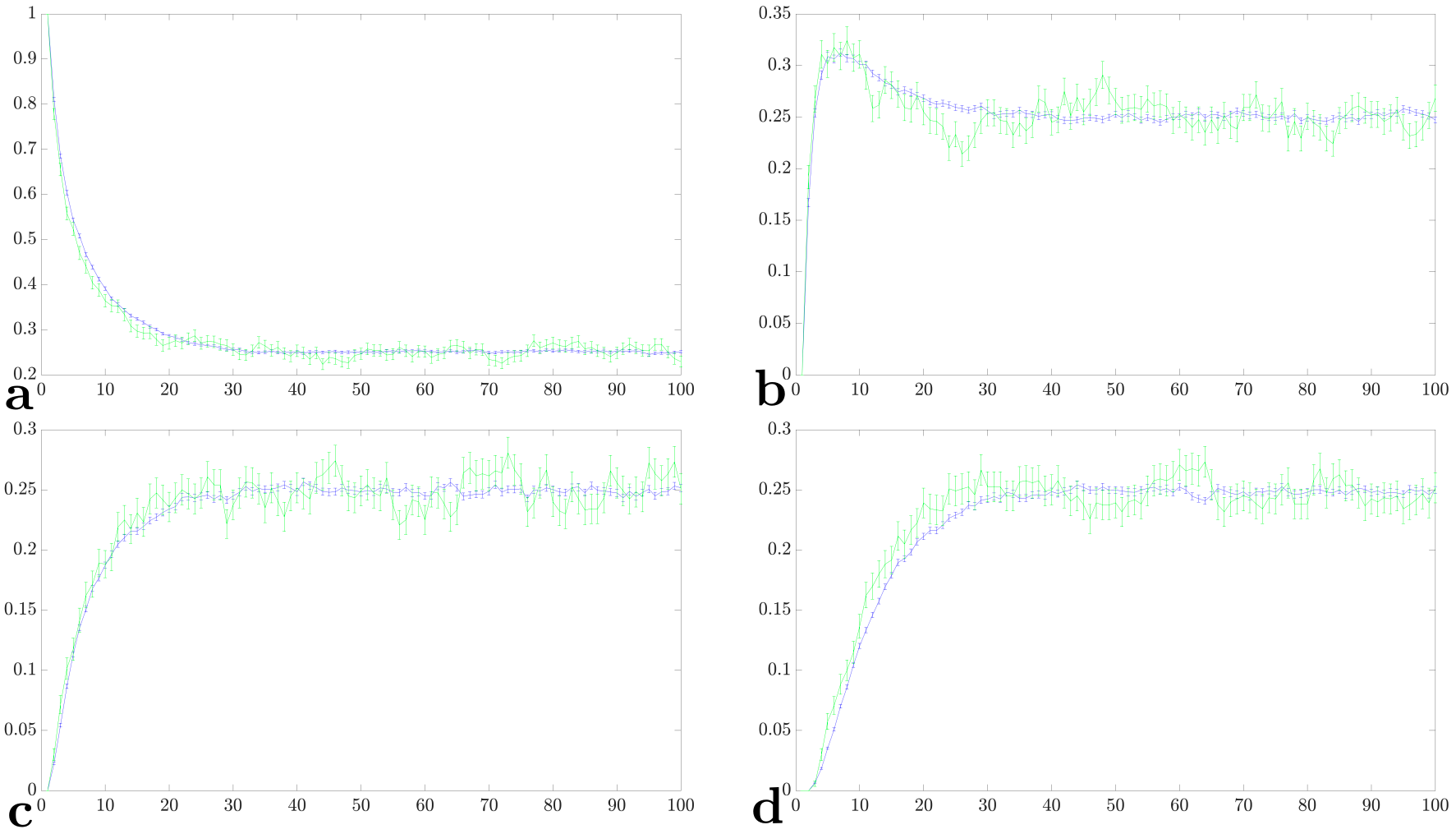}
\caption{This figure displays the results for the case where each site has an equal probability of being chosen for the entanglement and wave function collapse operations after each time step is completed. Each driving step is divided into 10 time steps. a) corresponds to the y-index of 0, b) corresponds to the y-index of 1, c) corresponds to the y-index of 2, and d) corresponds to the y-index of 3. The total number of noise realizations is 1200. The blue curve represents the case where the system is unacted upon by the entanglement or wave function collapse operations.}
\label{DarkEnergy10}
\end{figure*}

\begin{figure*}[t]
\centering
\includegraphics[width=1\textwidth]{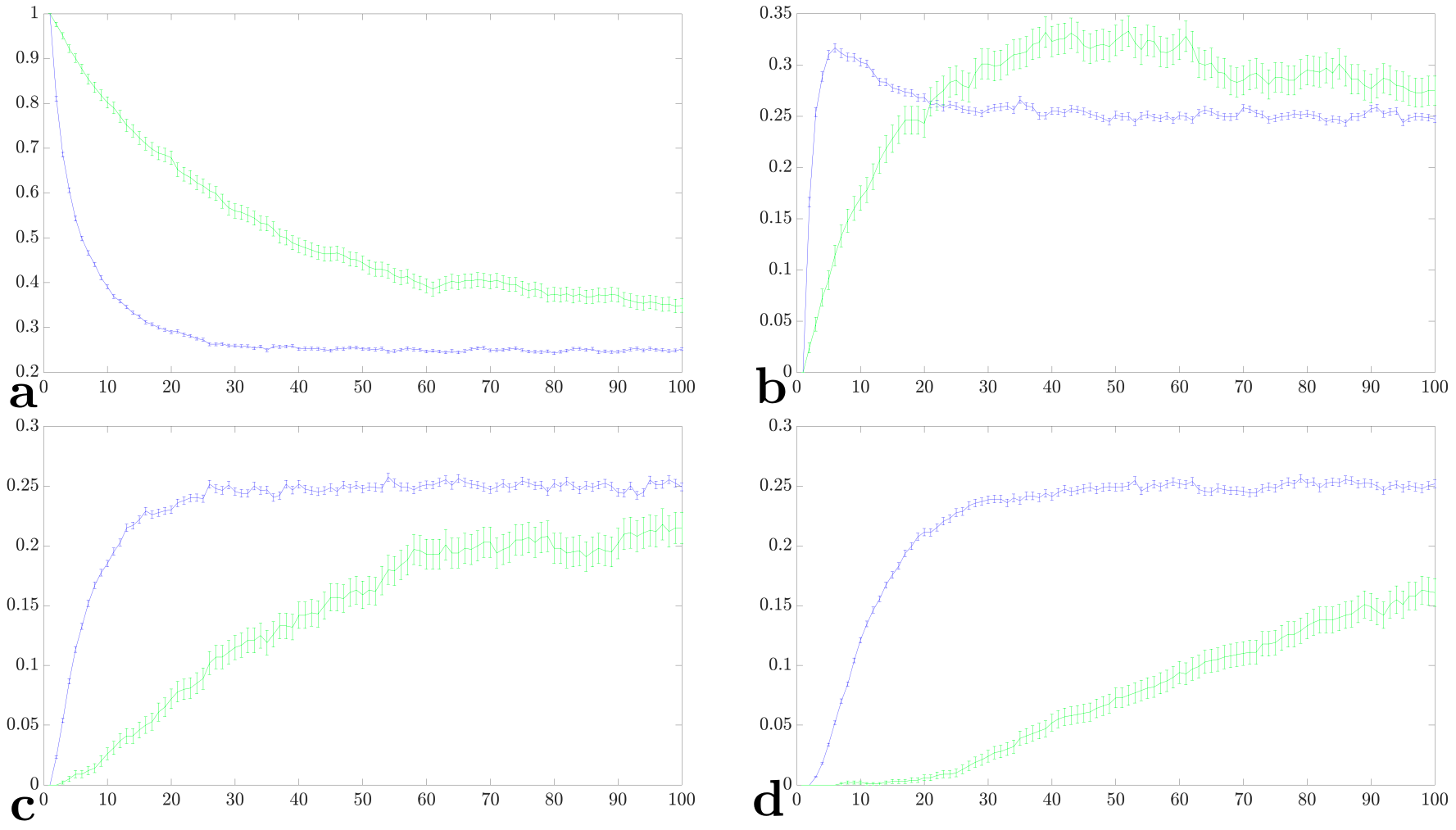}
\caption{This figure displays the results for the case where each site has an equal probability of being chosen for the entanglement and wave function collapse operations after each time step is completed. Each driving step is divided into 100 time steps. a) corresponds to the y-index of 0, b) corresponds to the y-index of 1, c) corresponds to the y-index of 2, and d) corresponds to the y-index of 3. The total number of noise realizations is 1000. The blue curve represents the case where the system is unacted upon by the entanglement or wave function collapse operations.}
\label{DarkEnergy100}
\end{figure*}

Figure~\ref{AntiGravityOccupation} displays the results when the model that uses wave function collapse and entanglement operations is implemented along with the dimensions $L_\mathrm{x}=2$ and $L_\mathrm{y}=8$. The probability vector for both the wave function collapse and entanglement operations is $p=[0.45\ (0.45+0.45/10)\ (0.45+0.45/10+0.45/100)\ 0.5\ (1 - 0.45-0.45/10-0.45/100)\ (1 - 0.45-0.45/10)\ (1 - 0.45)\ 1]$. The probability vector causes these operations to be least likely to be implemented for y-indices 3 and 4 and the probability steadily grows from y-indices 4 to 7 and also from 3 to 0. The probabilities (y-axis) for the particle to occupy each of the y-indices (x-axis) is obtained after the system has been evolved for 1000 driving cycles. Plots that display the probability of the particle occupying each of the y-indices for each driving step similar to Figures~\ref{Gravity},~\ref{DarkEnergy2},~\ref{DarkEnergy10}, and~\ref{DarkEnergy100} can be seen on this \href{https://github.com/htim327/QuantumGravity/blob/main/docs/source/AntiGravityPlots.rst}{GitHub page}.

The purpose of this setup is to generate a system where matter and energy `gravitates' away from the center and towards the edges, thereby producing an effective `negative gravitational mass' in the center~\cite{Bondi1957,Bonnor1989,Bonnor1964}. Figure~\ref{AntiGravityOccupation} clearly shows that this negative gravitational mass has been achieved according to the model presented. However, the figure also presents a problem with the existence of a standalone negative gravitational mass that exists in free space. In order for this negative gravitational mass to exist, the probability for the wave function collapse and entanglement operations would have to increase as the distance away from the particle increases. Even if the antigravitational force of this particle asymptotes to zero, the probability for these operations occuring would have to asymptote to a finite non-zero value at larger distances away from the particle. Therefore, a standalone negative gravitational mass clearly cannot be stable according to this model.

On the other hand, the problems associated with a standalone negative gravitational mass are not present for a local negative gravitational mass whose antigravitational fields are cancelled over longer distance by a positive gravitational mass(es) in the relative vicinity. This can be seen if the boundary conditions for the y-indices in Figure~\ref{AntiGravityOccupation} are periodic, such that y-index 7 is adjacent to y-index 0. In this case, a positive gravitational mass would be placed at y-indices 0 and 7, while a negative gravitational mass would be placed at y-indices 3 and 4 and the antigravitational fields would be cancelled at larger distances. It is unclear how stable this configuration would be without the use of external interactions executed by quantum information processing.

\section*{Discussion}

This study started out with a basic framework for how time dilation and length contraction can be modelled using the information stored on entangled wave functions within a quantum computer. But this still left some questions as to whether or not this setup could actually be used to simulate spacetime curvature within General Relativity. After a series of trials, a more well defined model was generated based on concepts similar to the original motivations that did, in fact, produce an effective force similar to that of gravity. This effective force relied on the usage of quantum entanglement as well as wave function collapse operations performed on the particle with these operations being more likely to occur for certain y-indices as opposed to others. The particle had a higher probability of being located at y-indices where these operations had a greater probability of occurring; thereby producing an effect similar to that of a gradient of the gravitational potential~\cite{Forshaw2014}. The required usage of wave function collapse operations complicates the relationship between the model found and the one that originally motivated this paper, but perhaps this problem can be resolved if the process of wave function collapse is dependent upon quantum entanglement operations~\cite{Wallace2012}. It must be reiterated, yet again, that the effective force found may not be the gravitational force, but this study gives arguments as to why this effective force is consistent with what is expected from a quantum gravity model.

This effective force relies on the transfer of quantum information and information transfer is one of the key concepts related to causality~\cite{Lizier2010}. When Special Relativity and General Relativity are considered as frameworks that describe how causal systems behave~\cite{Rahaman2021}, it is reasonable to assume that this effective force produced by the transfer of quantum information is in fact gravity and not just a system that behaves similar to or analogous to gravity. On top of this, the fact that this force requires the continuous transformation of the particle between a more de-localized (quantum) state and a more localized (classical) state and vice versa provides a further argument that this is a quantum gravitational effect that unites both the quantum and classical worlds~\cite{Wetterich2010,Rosen1986}.

The model also presents a mechanism through which singularities are prevented in environments such as black holes. Black holes are saturated with entropy because their entropy is by definition given by the Bekenstein bound~\cite{Bekenstein1981,Bekenstein2005,Bekenstein2008,Bousso2020}. In an environment where the black hole is unable to produce new entropy, such as when new matter and energy are not being deposited into the black hole, the `gravitational force' on the inside of the black hole would be similar to that seen in Figure~\ref{WaveFunctionCollapse}, which illustrates a `gravitational force' of zero. If matter and energy are in fact deposited into the black hole, the scenario admittedly becomes more complicated, but, in most cases, the amount of additional entropy produced is minuscule compared to the total entropy of the black hole~\cite{Onken2022}. It is unclear how the process of wave function collapse occurs for particles within a black hole due to the fact that the interior of a black hole cannot be observed and the measurement problem has not been solved~\cite{Mann2015,Hawking2010}.

Even if the effective force produced in this study turns out not to be the gravitational force, there are still a wide variety of uses for this effect. This effect can be utilized to orient random and chaotic systems to a more desired state or outcome. This can be potentially used to stabilize various phases of matter, such as superfluid or superconducting phases. It can also be potentially used to increase the probability of various events happening, such as the likelihood that a chemical or even fusion reaction occurs, the probability of a photon from the sun inducing a current within a photovoltaic cell, the probability of a set of qubits being initialized in the spin down state in a quantum computer, or for a substance to be transferred to a certain region.

It is beyond the scope of this study to determine how well the transfer of quantum information about various low energy states, such as those involving the hyperfine structure or low energy spin states, would equate into an effective force relative to information involving high energy states. For instance, if the goal is to have electrons occupy a particular orbital of an atom, then it might be prudent to perform the wave function collapse and entanglement operations using the hyperfine structure associated with those orbitals as opposed to the orbitals themselves. This could potentially be an effective tool to achieve this effective force without transferring very much heat to the particles of interest through noisy operations. This can be beneficial for larger scale operations where a high degree of accuracy for quantum operations are not possible~\cite{Tiurev2022}, such as concentrating carbon dioxide in a certain region so that it can be stored more efficiently~\cite{Leung2014} or to send atoms through a thruster without being ionized or oxidized~\cite{Kalentev2014,Trache2017}. In addition, it might be possible to perform the wave function collapse and entanglement operations on a low energy state of a reaction outcome as opposed to any low energy states that already exist within the system; thereby having the reactants `gravitate' towards a desired outcome. Many of these concepts might be fantasy for now, but only one concept is needed to demonstrate the usefulness of this topic.

The `gravitational' model was extended to explore concepts similar to dark energy where an evenly distributed energy field, which was modelled with the wave function collapse and entanglement operations having an equal probability of occurring throughout the system, causes matter and energy to diffuse more quickly. The setup was limited in terms of modelling how dark energy actually behaves in the universe, but did, in fact, observe quicker diffusion rates for low enough `dark energy' levels with higher `dark energy' levels causing slower diffusion rates in a process similar to time dilation~\cite{Thorne2000}. Finally, it was shown how this gravitational model can produce an effect similar to the existence of negative gravitational mass~\cite{Bondi1957,Bonnor1989,Bonnor1964}. This negative gravitational mass cannot exist as a standalone mass in free space, but must have a positive gravitational mass exist in the vicinity to cancel out the antigravitational fields at longer distances and thereby, cancel out various divergences. It is unclear how this negative gravitational mass can be used in the real world, if it can be created at all, but it would be totally unsurprising if a fundamental limit existed that prevented this negative gravitational mass from being used to generate mechanisms such as devices with over unity efficiency~\cite{Ford1978}.

\section*{Acknowledgements}

The author is grateful for the educational resources provided by The University of Texas at Dallas as well as those provided by the team at Amazon Braket. The Ganymede cluster operated by the University of Texas at Dallas' Cyberinfrastructure and Research Services Department was also used for computational resources.

\bibliographystyle{plain}
\bibliography{main}

\end{document}